\documentclass[11pt]{article}

\usepackage[utf8]{inputenc}
\usepackage[T1]{fontenc}
\usepackage{lmodern}
\usepackage[a4paper,margin=23mm]{geometry}
\usepackage{microtype}
\usepackage{booktabs}
\usepackage{tabularx}
\usepackage{array}
\usepackage{enumitem}
\usepackage{xcolor}
\usepackage{tikz}
\usetikzlibrary{arrows.meta,positioning,fit}
\usepackage[hidelinks]{hyperref}
\usepackage{fancyhdr}

\definecolor{scienceblue}{HTML}{174A67}
\definecolor{sciencegreen}{HTML}{34755B}
\definecolor{softblue}{HTML}{EAF3F7}
\definecolor{softgreen}{HTML}{ECF5F0}
\newcolumntype{Y}{>{\raggedright\arraybackslash}X}
\newcolumntype{P}[1]{>{\raggedright\arraybackslash\hyphenpenalty=10000\exhyphenpenalty=10000}p{#1}}
\setlist{leftmargin=*,itemsep=0.25em,topsep=0.35em}
\hypersetup{
  pdftitle={From Misconceptions to Evidence: What Science Teachers Make Visible When Co-Designing Agentic Learning Apps},
  pdfauthor={Nizam Kadir; Liow Wei Ting; Sumbul Khan; Ang Lay Kee},
  pdfsubject={Working paper; related abstract accepted for oral presentation at SISTC 2026},
  pdfkeywords={science teacher professional learning, educational AI, co-design, formative assessment, misconceptions, teacher agency, epistemic practices}
}

\title{\textbf{From Misconceptions to Evidence:}\\
What Science Teachers Make Visible When Co-Designing Agentic Learning Apps}
\author{Nizam Kadir \quad Liow Wei Ting \quad Sumbul Khan \quad Ang Lay Kee\\[0.6em]
\small Science, Mathematics and Technology (SMT), Singapore University of Technology and Design,\\[-0.1em]
\small Singapore 487372, Singapore}
\date{}

\begin{document}
\maketitle

\begin{abstract}
Science educators increasingly encounter AI tools that generate content, yet disciplinary teaching depends on eliciting learners' models, diagnosing misconceptions, interpreting evidence, and preserving professional judgment. This study asks how science teachers translate such epistemic work into specifications for agentic learning applications. It contributes to the conference theme, ``Innovating Pedagogies, Inspiring Minds: Transforming Science Learning,'' and the Teachers' Professional Learning strand by examining app co-design as a form of pedagogical reasoning. We conducted a bounded qualitative cross-case analysis of four de-identified artifacts produced in a teacher professional-learning workshop: an experimental-design diagnostic, a Kinetic Particle Theory dialogue guide, a chemistry prior-knowledge checker, and a physics application/scaffolding tool. Each artifact was coded for the disciplinary problem, learner interaction, evidence made visible, teacher authority, and safeguard.

All four connected a science-learning problem to an interaction and pedagogically interpretable evidence: misconceptions and gaps, explanations-in-progress, class-level readiness patterns, or investigation performance. However, only two made teacher control or evaluation explicit, and only two named a safeguard. The proposals therefore positioned AI less as an answer generator than as an elicitor, scaffold, and evidence-return mechanism, while leaving decision rights and protections unevenly specified.

We argue that teacher professional learning should treat AI app ideation as epistemic specification work. A five-question design protocol---problem, learner interaction, evidence, teacher authority, and safeguard---can help teachers transform science-learning needs into accountable human--AI arrangements before building or adopting a tool.
\end{abstract}


\begin{center}
\small\itshape
The related abstract with the same title has been accepted for a 25-minute oral presentation followed by 10 minutes of Q\&A at the 8th Singapore International Science Teachers' Conference (SISTC 2026), Science Centre Singapore, 24--26 November 2026. This full working paper has not been peer reviewed or accepted for conference proceedings.
\end{center}

\section{Introduction}

Generative artificial intelligence has made it possible to imagine educational software from a short natural-language description. For science teachers, that promise is attractive: a locally meaningful difficulty might become an interactive diagnostic, a dialogue guide, or a scaffolded investigation without waiting for a commercial product to address the same context. Yet science teaching is not a generic content-delivery problem. Teachers decide which misconception matters, what learners should explain or investigate, what counts as evidence of understanding, and when feedback must stop short of supplying an answer. A useful application must embody these judgments, not simply reproduce the topic vocabulary of a prompt.

This distinction matters because science learning involves conceptual, epistemic, and social goals \cite{duschl2008}. Learners do not only acquire facts; they build and revise models, coordinate claims with evidence, design investigations, and make phenomena intelligible. Formative assessment likewise depends on eliciting student thinking, interpreting it against a learning goal, and deciding what action follows \cite{black1998,ruizprimo2007}. An AI application can support this work only if its proposed interaction preserves the relationship among learner action, evidence, and professional interpretation.

Teacher co-design offers a way to examine that relationship before software is built. Collaborative design can surface professional knowledge that is otherwise difficult to encode: classroom routines, disciplinary bottlenecks, usable evidence, and boundaries on automation \cite{mckenney2016}. App ideation is therefore not merely brainstorming. When teachers specify what an application should ask, notice, return, and leave to human judgment, they externalize a provisional theory of teaching and learning.

Existing work on educational AI has emphasized teacher--AI complementarity, classroom orchestration, and shared control \cite{holstein2019,molenaar2022,lawrence2024}. Less attention has been paid to the small but consequential design artifacts through which teachers translate a disciplinary problem into an imagined agentic interaction. We use \emph{agentic learning application} as a working term for an application that delegates a responsive action---such as questioning, diagnosis, scaffolding, or feedback---while learners and teachers remain part of the activity. The term describes a proposed interaction, not evidence that an autonomous system was implemented.

We analyze four de-identified science-specific artifacts from a teacher professional-learning workshop. The cases concern experimental design, Kinetic Particle Theory, chemistry prerequisite knowledge, and the application of physics to real-world situations. Their small number supports depth and contrast, not prevalence claims. We ask:

\begin{enumerate}[label=\textbf{RQ\arabic*.}]
\item How did four science-specific workshop artifacts translate disciplinary problems of practice into proposed agentic learning interactions?
\item What forms of pedagogical action, evidence, teacher authority, and safeguards became visible---and which remained underspecified---in those artifacts?
\end{enumerate}

The analysis makes two contributions. Empirically, it shows that all four artifacts organized AI around the elicitation, scaffolding, or interpretation of scientific reasoning, and made some form of pedagogical evidence visible. Analytically, it identifies an \emph{epistemic specification} problem: learner interaction and evidence were consistently articulated, while teacher decision rights and safeguards were explicit in only half of the cases. We translate this result into a five-question protocol for science-teacher professional learning. The contribution is deliberately bounded: these artifacts reveal what was specified during co-design, not whether any proposed application would work or improve learning.

\section{Conceptual Background}

\subsection{Science learning as epistemic work}

Contemporary accounts of science education treat scientific practices as inseparable from disciplinary knowledge. Learners are expected to ask questions, develop and use models, plan investigations, analyze data, construct explanations, and argue from evidence \cite{nrc2012}. These practices are epistemic because they concern how knowledge claims are made, tested, revised, and justified. Duschl's account of science education in ``three-part harmony'' similarly connects conceptual structures with epistemic practices and social processes \cite{duschl2008}. A learning tool that produces a correct answer can therefore miss the activity that makes the answer educationally meaningful.

Misconceptions and incomplete models are especially revealing design objects. They are not simply errors to erase. They provide evidence about how a learner is coordinating concepts and representations. In Kinetic Particle Theory, for example, learners must reason about an invisible model and connect it to observable phenomena. In experimental design, the challenge is not only naming variables but coordinating a question, procedure, evidence, and warranted conclusion. A pedagogically useful interaction must make these relations inspectable enough for learners to revise and teachers to interpret.

This perspective changes what ``evidence'' means in educational software. Clicks, completion, and time-on-task can describe activity without showing the quality of a model, explanation, or investigation. Evidence for science learning is more closely tied to what students say, represent, revise, or do. An application specification therefore carries an epistemic commitment whenever it defines the trace that returns to a learner or teacher.

\subsection{Formative assessment as an evidence-to-action relation}

Formative assessment is often described as the use of evidence to adapt teaching and learning \cite{black1998}. In science classrooms, informal formative assessment unfolds through linked moves: a teacher elicits thinking, recognizes its significance, and uses it to advance understanding \cite{ruizprimo2007}. The evidence has value because someone can interpret it and act. A list of detected misconceptions, for example, is not yet formative unless it informs a learner's revision or a teacher's next decision.

This evidence-to-action relation provides a useful lens for agentic applications. A diagnostic agent might pose questions and infer gaps, but its educational role depends on the questions' disciplinary meaning, the feedback boundary, the visibility of the learner's reasoning, and the authority to interpret or override an inference. The same logic applies to dialogic guidance and scaffolded investigation. Automation can widen the opportunity to elicit evidence, but it can also obscure how an inference was made or encourage a premature answer. The design challenge is not maximizing automation; it is specifying a defensible relation among learner action, machine response, evidence, and teacher judgment.

\subsection{Teacher agency, co-design, and educational AI}

Teacher agency involves more than approving a machine recommendation. It includes shaping goals, interpreting context, selecting resources, and acting responsibly within institutional conditions \cite{imants2020}. Research on teacher--AI complementarity accordingly emphasizes tools that make analyses actionable within classroom orchestration and preserve opportunities for professional intervention \cite{holstein2019}. Work on hybrid human--AI learning technologies and shared-control systems likewise frames authority as a negotiated arrangement rather than a binary choice between human and machine \cite{molenaar2022,lawrence2024}.

Co-design can make these arrangements discussable. When teachers formulate an application, they decide what the learner does, what an automated component may do, and what information returns. They may also state who selects content, reviews an output, grades performance, or protects privacy. These elements constitute an educational specification even when expressed in ordinary language. We call it an \emph{epistemic specification} when it links a disciplinary problem to a learner interaction, a form of evidence, decision rights, and safeguards. This lens treats professional learning as an opportunity to make tacit pedagogical judgments explicit and contestable before technology is adopted.

\section{Method}

\subsection{Design and corpus}

We conducted a bounded qualitative cross-case analysis of four de-identified artifacts produced during a science-teacher professional-learning and co-design activity. The activity asked teachers to begin with a problem of practice and imagine an agentic learning application. One artifact was posted as a group framing and three were later structured proposals. We treat each artifact as a design expression rather than as an individual participant response because group reporting and individual authorship cannot be equated.

Cases were included when the preserved artifact (1) named a science-specific disciplinary problem, and (2) described a responsive application interaction. The resulting corpus comprised:

\begin{itemize}
\item an experimental-design diagnostic using three questions, feedback, and reflection (S1);
\item a Kinetic Particle Theory guide using dialogic questioning and differentiated guidance (S2);
\item a chemistry prior-knowledge checker focused on formulas and equations required before mole calculations (S3); and
\item a physics application/scaffolding proposal using real-world scenarios and student video investigations (S4).
\end{itemize}

No implementation, classroom-use, or outcome data are part of this study. The unit of analysis is the content made explicit in each artifact.

\subsection{Analytic procedure}

Each case was summarized in a structured matrix using five dimensions derived from formative assessment, science practices, and teacher-agency literature:

\begin{enumerate}
\item \textbf{Disciplinary problem:} the science-learning difficulty or epistemic bottleneck;
\item \textbf{Learner interaction:} what the learner would answer, explain, revise, or investigate, and how the application would respond;
\item \textbf{Evidence:} the trace made available for learner or teacher interpretation;
\item \textbf{Teacher authority:} explicit rights to select, review, edit, evaluate, grade, or release; and
\item \textbf{Safeguard:} an explicit boundary concerning privacy, human review, accuracy, fairness, or dependence.
\end{enumerate}

A dimension was marked present only when the artifact stated or directly entailed it. We did not infer a teacher-control mechanism from the educational setting alone, or treat generic data collection as pedagogical evidence. We first wrote within-case summaries, then compared the cases for common relations and negative contrasts. Counts are used descriptively to make the coding transparent; with four artifacts, they are not statistical estimates.

\begin{table}[htbp]
\centering
\caption{Four-case science specification matrix. ``Not explicit'' describes the preserved artifact, not the teacher's unexpressed beliefs.}
\label{tab:cases}
\small
\begin{tabularx}{\textwidth}{@{}P{0.55cm}Y Y Y Y@{}}
\toprule
\textbf{ID} & \textbf{Epistemic problem} & \textbf{Proposed interaction} & \textbf{Evidence made visible} & \textbf{Authority and safeguard} \\
\midrule
S1 & Gaps or misconceptions in experimental design & Three-question diagnostic, personalized feedback, and reflection & Diagnosed gaps and misconceptions & Neither explicit \\
S2 & Invisible Kinetic Particle Theory models are difficult to explain & Dialogic questioning, differentiated guidance, and answer formulation & Teacher-facing account of conceptual shortcomings & Privacy stated; no direct approval/edit right \\
S3 & Missing formulas and equations before mole calculations & Short diagnostic followed by misconception-sensitive repair & Class-level gaps and proportion needing support & Teacher selects assessed concepts; no safeguard explicit \\
S4 & Difficulty applying physics in real-world situations & Scenarios, scaffolded knowledge building, and video investigations & Investigation performance for teacher evaluation & Teacher grades/evaluates; human review stated \\
\bottomrule
\end{tabularx}
\end{table}

\section{Findings}

\subsection{The cases began with epistemic bottlenecks, not content topics alone}

Each artifact named a difficulty in how learners know or practice science. S1 focused on gaps and misconceptions in the design of experiments. Its object was not an experiment-themed quiz in general, but the reasoning needed to make an investigation coherent. S2 addressed the difficulty of explaining an invisible particle model, where a learner must connect a representation to observable behavior. S3 placed prerequisite knowledge---formulas and equations---before mole calculations, treating later performance as dependent on an interpretable readiness structure. S4 addressed the transfer of physics knowledge into real-world situations and investigations.

This commonality is consequential. The proposed applications were not organized around delivering more information about experiments, particles, moles, or physics. They were organized around points where learners' thinking could become visible: a design decision, a developing explanation, a pattern of prerequisite gaps, or an investigation performance. The four artifacts therefore translated curriculum content into a candidate epistemic interaction.

The cases also differed in the grain of the problem. S1 and S3 framed a diagnostic moment: determine what the learner knows before or during further instruction. S2 framed a dialogic process in which an explanation develops through questioning. S4 framed a performance sequence in which knowledge is scaffolded toward an observable investigation. This variation shows that ``agentic app'' did not imply one interface genre. The common unit was responsive action around learner reasoning.

\subsection{Proposed agency centered elicitation, scaffolding, and repair}

All four artifacts specified what an application would do in response to learners. S1 compressed diagnosis into three questions, followed by personalized feedback and reflection. S2 proposed dialogic questioning and differentiated guidance that would support learners in formulating an answer. S3 paired a short diagnostic with misconception-sensitive repair. S4 combined contextual scenarios, scaffolded knowledge building, and video investigations.

Across the set, the system's role was bounded but active. It would elicit, question, diagnose, differentiate, scaffold, or feed back. None of the four was centered on producing a finished scientific answer for the learner. This does not prove that an implementation would preserve productive struggle; it does show that the proposed value of agency lay in adapting a pathway around evidence of thinking.

The interaction structures also imply different feedback timings. S1 and S3 envisioned relatively short diagnostic cycles. S2 implied a multi-turn dialogue in which guidance changes as an explanation develops. S4 implied an extended sequence from scenario to knowledge construction to documented investigation. Professional learning about agentic tools should therefore move beyond the question ``What should the AI generate?'' to questions about the tempo, stopping conditions, and evidence transitions of the learning activity.

\subsection{Every artifact specified evidence with disciplinary meaning}

All four artifacts made a form of evidence visible (Table~\ref{tab:cases}). S1 would return diagnosed gaps and misconceptions. S2 would make conceptual shortcomings available to the teacher while learners formulated explanations. S3 would aggregate class-level gaps and the proportion of learners needing support. S4 would preserve investigation performance in a form that a teacher could evaluate.

These traces differ from generic engagement telemetry. Their value comes from their relation to a disciplinary judgment: whether an experimental design is coherent, whether a particle explanation is developing, whether prerequisite chemistry knowledge is present, or whether physics knowledge can be applied. The cases thus frame evidence as a bridge between learner action and pedagogical response.

They also distribute evidence differently. S1 appears oriented toward feedback and learner reflection; S2 explicitly returns shortcomings to the teacher; S3 produces a class-level readiness view; and S4 produces an assessable performance artifact. The envisioned audience shifts from learner to teacher and from individual to class. A specification must therefore name not only what data exist but who can interpret them, at what level of aggregation, and for which next action.

\subsection{Authority and safeguards were unevenly specified}

The strongest cross-case contrast concerns governance. While problem, interaction, and evidence were visible in all four artifacts, explicit teacher authority appeared in two and an explicit safeguard appeared in two (Table~\ref{tab:coverage}). S3 gave the teacher authority to select which concepts the diagnostic would assess. S4 reserved evaluation or grading for the teacher and named human review. S2 stated a privacy safeguard but did not specify a direct teacher approval or editing right. S1 did not make either dimension explicit.

\begin{table}[htbp]
\centering
\caption{Descriptive coverage of the five analytic dimensions ($N=4$ artifacts).}
\label{tab:coverage}
\begin{tabular}{@{}lr@{}}
\toprule
\textbf{Dimension made explicit} & \textbf{Artifacts} \\
\midrule
Disciplinary problem & 4/4 \\
Learner interaction & 4/4 \\
Pedagogically interpretable evidence & 4/4 \\
Teacher authority or evaluation & 2/4 \\
Safeguard & 2/4 \\
\bottomrule
\end{tabular}
\end{table}

The pattern is not evidence that teachers considered authority or safety unimportant. A short artifact can omit what its author assumes, and the elicitation format may give more space to functionality than governance. The defensible result is narrower: evidence and interaction were consistently legible in the preserved specifications, whereas decision rights and protections were not consistently legible. For anyone reviewing, building, or adopting the proposed applications, those omissions would leave consequential questions unresolved.

The contrast between S2 and S3 is particularly instructive. S2 named privacy but not who could approve or edit the system's differentiated guidance. S3 gave teachers control over diagnostic content but named no safeguard for the resulting misconception data. Safeguards and authority are therefore not interchangeable. Privacy without decision rights can leave feedback ungoverned; teacher configuration without a data boundary can leave evidence use underspecified.

\section{Discussion}

\subsection{App co-design externalized science teachers' epistemic judgments}

The four artifacts support an interpretation of co-design as epistemic specification work. Teachers translated curricular concerns into relations among a learner, a responsive mechanism, and evidence. In doing so, they made aspects of professional judgment visible that a feature-first conversation could obscure. ``Three diagnostic questions,'' ``dialogic questioning,'' ``class-level gaps,'' and ``video investigations'' each define more than functionality. They imply a theory about where understanding becomes observable and what kind of response might move it forward.

This interpretation connects science practices with teacher professional learning. If science teaching involves coordinating conceptual, epistemic, and social goals \cite{duschl2008}, then designing an AI-supported activity can become an occasion to articulate that coordination. A teacher must ask which reasoning should remain with the learner, what the system may infer, and what evidence deserves professional attention. Co-design can thus shift AI professional learning away from prompt tricks or product demonstrations toward examination of pedagogical commitments.

The result also refines the meaning of personalization. In these cases, personalization was not primarily content matching. It involved responding to an experimental-design gap, differentiating questions around an invisible model, repairing prerequisite misconceptions, or scaffolding transfer to a situation. Personalization becomes educationally meaningful when it is anchored in a disciplinary model of learner thinking and produces evidence that supports a next action.

\subsection{Evidence should be specified as part of an instructional relation}

All four artifacts made evidence central, but the evidence varied in audience, grain, and temporal role. This finding suggests that app-design activities should separate at least three questions: What learner action produces the evidence? What disciplinary interpretation can be made from it? What decision or revision follows? Without this chain, a dashboard can make data visible without making learning intelligible.

The distinction matters for student learning. A misconception label may help if it prompts a learner to revise an explanation or helps a teacher select an intervention. It may harm if it becomes a fixed classification or if the inference is opaque. A video investigation may reveal application and reasoning, but it also raises questions about privacy, access, assessment criteria, and review workload. Evidence is therefore not an unqualified benefit. Its pedagogical value and ethical burden emerge together.

Formative assessment research emphasizes that evidence becomes formative through use \cite{black1998,ruizprimo2007}. Our cases extend that principle to app specification: the evidence field should be connected explicitly to an interpretive actor and a next instructional move. This requirement is especially important when an automated component may infer misconceptions or personalize feedback.

\subsection{Epistemic visibility without governance is incomplete}

The uneven coverage of authority and safeguards exposes an accountability gap. Making student thinking visible can expand teachers' capacity to respond, but it can also increase surveillance, misclassification, or professional workload. Teacher oversight cannot be added as a generic ``human in the loop'' statement after an interaction has been designed. It must identify the decision: selecting concepts, editing feedback, reviewing an inference, grading evidence, releasing information, or handling an exception.

Similarly, a safeguard should respond to the actual evidence flow. Privacy was explicit in S2, where conceptual shortcomings returned to a teacher. Human review was explicit in S4, where video investigations supported evaluation. Other possible safeguards---data minimization, transparent inference, learner contestability, bounded retention, and accessibility---were not made explicit in this small corpus and should not be attributed to it. Their absence from the artifacts identifies questions for facilitation rather than faults in an implemented system.

Teacher--AI complementarity depends on shared control that is tied to classroom practice \cite{holstein2019,lawrence2024}. The present analysis suggests that such control can begin at ideation. Before a tool exists, teachers can specify what an automated component may do, what evidence it may return, and which judgments remain professional. This makes governance part of pedagogical design rather than a compliance addendum.

\section{Implications for Science Teaching and Teacher Learning}

\begin{figure}[htbp]
\centering
\begin{tikzpicture}[
  node distance=8mm and 7mm,
  box/.style={rounded corners=2pt, draw=scienceblue, very thick, fill=softblue, align=center, text width=3.45cm, minimum height=1.35cm, inner sep=6pt},
  gov/.style={rounded corners=2pt, draw=sciencegreen, very thick, fill=softgreen, align=center, text width=5.15cm, minimum height=1.15cm, inner sep=6pt},
  arrow/.style={-{Latex[length=2.5mm]}, very thick, draw=scienceblue}
]
\node[box] (problem) {\textbf{1. Problem}\\Which disciplinary difficulty or misconception?};
\node[box, right=of problem] (interaction) {\textbf{2. Learner interaction}\\What must the learner explain, revise, or investigate?};
\node[box, right=of interaction] (evidence) {\textbf{3. Evidence}\\What trace supports interpretation and a next action?};
\draw[arrow] (problem) -- (interaction);
\draw[arrow] (interaction) -- (evidence);
\node[gov, below=10mm of problem, xshift=28mm] (authority) {\textbf{4. Teacher authority}\\What must a teacher select, review, evaluate, or release?};
\node[gov, right=8mm of authority] (safeguard) {\textbf{5. Safeguard}\\What privacy, accuracy, fairness, dependence, or human-review boundary?};
\node[draw=sciencegreen, rounded corners=3pt, inner sep=5pt, fit=(authority)(safeguard), label={[text=sciencegreen]below:{Govern the complete evidence-to-action relation}}] {};
\end{tikzpicture}
\caption{The five-question epistemic specification protocol. The figure is a conceptual synthesis from the four-case analysis, not an evaluated intervention.}
\label{fig:protocol}
\end{figure}
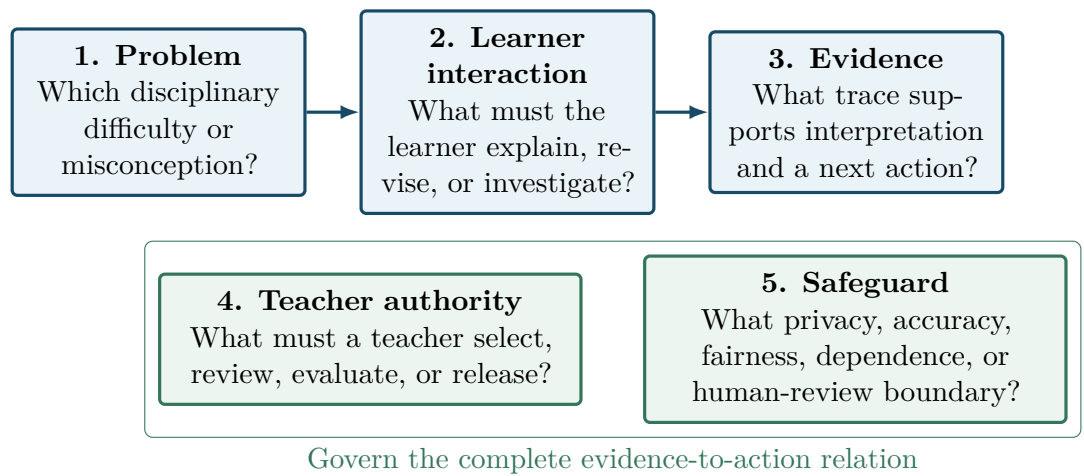

The findings motivate a five-question specification protocol (Figure~\ref{fig:protocol}). The first three questions make the epistemic sequence visible; the final two govern every transition.

\subsection{For teacher professional learning}

Professional-learning facilitators can use the protocol in three passes. First, teachers describe a persistent science-learning problem without naming an AI feature. Second, they draw the evidence-to-action relation: learner action, responsive support, evidence, interpretation, and next move. Third, peers conduct an authority-and-safeguard review. A specification is ready for further design only when another teacher can identify who decides, what is visible, and what boundary prevents an unacceptable use.

The protocol can be applied with paper, slides, or a shared document; it does not require access to an app builder. This is important because the professional-learning objective is not rapid production. It is disciplined anticipation. Comparing specifications across topics can also expose how the same feature label---for example, ``personalized feedback''---means different things in experimental design, particle models, stoichiometric readiness, and physics investigations.

\subsection{For science teaching and prospective student learning}

For science teaching, the protocol encourages tools that preserve productive epistemic work. A diagnostic should reveal reasoning rather than only correctness. Dialogic guidance should help a learner formulate and revise an explanation rather than converge immediately on a supplied answer. Class-level summaries should remain traceable to meaningful concepts, and performance evidence should be evaluated with visible criteria.

These are design implications, not measured student outcomes. The four artifacts suggest plausible ways to support student learning by eliciting models, targeting misconceptions, and connecting evidence to feedback. Whether such designs improve understanding would require implementation, classroom enactment, and outcome measures. A next-stage study could compare specifications produced with and without the five-question protocol, then examine whether implemented interactions preserve the intended learner work and teacher decision rights.

\section{Limitations}

This study analyzes four artifacts from one professional-learning setting. The cases were selected for science specificity and analyzable interaction content, so they cannot represent the full workshop or science teachers generally. One artifact was group-authored and the others were posted proposals; artifact counts are not participant counts. The preserved text does not capture all spoken deliberation, assumptions, or revisions that may have occurred.

The analysis examines what was made explicit, not what teachers knew or valued privately. An omitted safeguard may reflect brevity, elicitation order, or tacit shared norms rather than lack of concern. The coding framework also privileges five dimensions chosen from prior literature; other readings could foreground curriculum alignment, accessibility, assessment validity, or classroom feasibility.

Most importantly, the artifacts describe prospective applications. No claim can be made about implementation fidelity, technical feasibility, usability, teacher workload, student experience, or learning effects. The proposed protocol is a conceptual implication and has not been evaluated. Future work should retain provenance from problem statement through specification and enactment, invite teachers to review analytic interpretations, and examine how students experience the resulting evidence and feedback relations.

\section{Conclusion}

Four science-teacher design artifacts reveal a consistent orientation: agentic learning applications were imagined as ways to elicit, scaffold, diagnose, and make scientific reasoning visible. Experimental design, particle models, prerequisite chemistry knowledge, and real-world physics application became different evidence-to-action relations rather than generic requests for content generation. Yet the specifications made teacher authority and safeguards explicit less consistently than they made learner interaction and evidence explicit.

The resulting professional-learning opportunity is to treat AI app ideation as epistemic specification work. Asking about the problem, learner interaction, evidence, authority, and safeguard can help teachers articulate what a proposed application should make visible and what it must leave under professional control. Such articulation does not guarantee a good tool, but it creates a stronger object for critique before design decisions become software and before software enters a science classroom.

\section*{Acknowledgement}

This work acknowledges funding support from the Design-centric Interdisciplinary Creative Problem Solving (ICPS-D): Developing Critical and Creative Skills in Teachers project.

This work was supported by the AIEDx grant from the Institute of AI in Education (IAIED), Singapore (UEN 53487769M).

\end{document}